# High Harmonic Radiation from a Low-$K$ Biharmonic Planar Undulator

Sihong Liu [a], Bocheng Jiang [b,*]

[a] College of Physics, Chongqing University, Chongqing, 401331, China

[b] Laboratory for Ultrafast Transient Facility, Chongqing University, Chongqing, 400044, Chian, Email: jiangbocheng@cqu.edu.cn

**Abstract**

High harmonic generation by an undulator is a key issue for extending the photon energy range of synchrotron light sources. In this work, we propose a biharmonic planar undulator operating in the low-$K$ regime ($K<1$) to enhance high-harmonic radiation. By superimposing a 1/3 subharmonic undulator field onto a short-period superconducting undulator, a biharmonic undulator is formed. The on-axis intensities of the 5th and 7th harmonics are significantly enhanced, corresponding to radiation near the 2nd harmonic of the short-period superconducting undulator. The results are confirmed by both theoretical analysis and numerical simulations using SPECTRA. It shows that the biharmonic configuration effectively overcomes the limitations of low-$K$ undulators for high harmonic generation. In this study, reasonable undulator parameters are selected with full consideration of technical challenges. The simulation indicates that an unprecedented photon flux in the 16–22 keV energy range can be achieved with a 3.5 GeV beam, offering a promising approach for high brightness, short-wavelength synchrotron radiation.

## 1. Introduction

Producing high flux short-wavelength radiation in synchrotron light sources represents one of the most important directions in the development of accelerator technology. The radiation wavelength is inversely proportional to $\gamma^2$ (where γ is the relativistic energy factor of the beam), Increasing the beam energy is a direct approach, but it is costly. Undulators are widely used in 3rd and 4th generation light sources and are indispensable for producing high flux radiation within a narrow bandwidth. The key mechanism is that radiation from different undulator poles interferes constructively under the resonance condition, and the radiated power at the resonant frequency scales as $N^2$, where $N$ is the number of periods.

The development of various types of undulators to meet user demands is an ongoing theme in synchrotron radiation research. Among these, the development of short-period undulators is a major direction. Compared with increasing the beam energy, shorting the magnetic period length of the undulator is much less costly to decrease the radiation wavelength.

To date, superconducting undulators represent a key technology for achieving magnetic period lengths below 18 mm. Even for superconducting undulators, the performance improvement remains relatively limited. This is because the development of short-period undulators faces several challenges. As the period length decreases, the winding space for superconducting wire becomes increasingly constrained. With the critical current of the superconducting wire fixed, the achievable magnetic field tends to decrease. However, in order to obtain high radiation flux, particularly for higher harmonics, the peak magnetic field must be increased to prevent a significant reduction of the undulator parameter $K$,

$$K = 0.934B(T)\lambda_u(cm),\tag{1}$$

where $B$ is the peak magnet field, $\lambda_u$ is the undulator period length.

Furthermore, the peak magnetic field scales with the ratio $Exp[-\pi(g/\lambda_u - 0.5)]$ ( $g/\lambda_u$ is undulator gap to period length)[1], achieving a high field in short-period undulators requires reducing the magnetic gap. In practice, the gap reduction is often limited by operational stability requirements. Consequently, further increasing the magnetic field under these constraints becomes extremely challenging.

A worldwide overview of progress in superconducting undulators is given in Ref. [2]. The period length of beam commissioned devices is typically in the range of 14-16 mm, which is not significantly shorter than that of in-vacuum permanent-magnet undulators (~20 mm), nor than that of cryogenic permanent-magnet undulators which can achieve period lengths of about 18 mm [3]. Numerous efforts have been devoted to developing new types of short-period undulators, including, for example, superconducting bulk array undulators [4], RF undulators [5] and laser undulators [6,7]. However, none of these concepts has yet reached the stage of practical application.

The record field achieved by superconducting bulk array undulators is 2.1 T with a period length of 10 mm [8], the $K$ parameter can reach 1.96. However, a couple of issues still need to be addressed for its feasibility. Since this scheme operates in a passive manner without a power supply, the magnetic field is generated by an external superconducting solenoid. As a result, the magnetic field may decay, or the bulk may quench due to the radiation environment or a failure of the cryogenic system. In fact, field decay has already been observed even in the absence of beam tests [8]. This issue deserves careful consideration since re-magnetizing the superconducting bulk is a time consuming task and may even require breaking the vacuum. Another concern is that the strong longitudinal magnetic field may lead to complex beam dynamics as particles traverse the device, which requires further investigation.

For a conventional superconducting undulator, according to the scaling law [1], the peak magnetic field is about 0.68 T for a period length of 8 mm with $g/\lambda_u = 0.5$. Under these conditions, the undulator parameter $K$ is about 0.5, for which, high harmonics can hardly be generated.

Exploring new methods to generate high harmonics in the low-$K$ regime is of considerable importance. To enhance high harmonic radiation, a biharmonic planar undulator is proposed in this paper. Biharmonic planar undulators have been previously studied in Refs. [9-12], the analytical results have been achieved both for helical and planar undulator with harmonic field. The major application for such undulator is to produce free electron laser in previous study. Here, we focus on the regime of $K<1$ to investigate how a biharmonic planar undulator can enhance high harmonic radiation which is exactly the technical challenge confronting short-period superconducting undulators at present.

2. **Biharmonic planar undulator theory**

In this section, we briefly review the theoretical framework presented in Ref. [11,12]. A biharmonic planar undulator with the on-axis magnetic field given by Eq. (2).

$$B_y = B_0[\sin(k_u z) + d\sin(hk_u z)], \quad h \equiv integer \tag{2}$$

Where $h$ is harmonic number of the undulator periods, $d$ is the magnetic field coefficient. The on-axis brightness of the radiation can be obtained by the following Eqs.

$$\frac{d^2I}{d\omega d\Omega} = \frac{e^2 N^2 \gamma^2}{4c} \sum_n \left[\frac{\sin(v_n/2)}{(v_n/2)}\right]^2 S_n \tag{3}$$

Here, $n$ is the harmonic number of the radiation, $c$ is the speed of the light, $e$ is the charge of electron.

$$S_n = \frac{n^2 K^2}{\{[1+(K^2/2)][1+(d/h)^2]\}^2} \left\{ [T_{n-1}(n\ arg) + T_{n+1}(n\ arg)] + \frac{d}{h}[T_{n+h}(n\ arg) + T_{n-h}(n\ arg)] \right\}^2 \tag{4}$$

$$T_n(arg) = \frac{1}{2\pi} \int_0^{2\pi} \cos[n\phi + \xi \sin(2\phi) + \xi_- \sin[(h-1)\phi] + \xi_+ \sin[(h+1)\phi] + \xi_h \sin(2h\phi)]\, d\phi \tag{5}$$

$$\nu_n = 2\pi n N \frac{\omega - n\omega_R}{n\omega_R} \tag{6}$$

$$\omega_R = \frac{2\pi c}{\lambda_R},\ \lambda_R = \frac{\lambda_u}{2\gamma^2}\left\{1 + \frac{K^2}{2}\left[1 + \left(\frac{d}{h}\right)^2\right]\right\} \tag{7}$$

$$\xi = \frac{1}{4}\frac{K^2}{1+\left(\frac{K^2}{2}\right)\left[1+\left(\frac{d}{h}\right)^2\right]},\ \xi_- = \frac{4d}{h(h-1)}\xi,\ \xi_+ = \frac{4d}{h(h+1)}\xi,\ \xi_h = \frac{d^2}{h^3}\xi \tag{8}$$

We first investigate whether high harmonic radiation can be generated by adjusting the subharmonic magnetic field while keeping the short-period magnetic field fixed. For the following study, we set $B_0 d \equiv 0.68\ T, hk_u \equiv 2\pi/8mm$. We then consider the superposition of a 1/3 subharmonic undulator with a period length of 24 mm, for which the peak magnetic field is varied from 0.1 T to 0.8 T. The on-axis radiation intensity is proportional to $S_n$, which is calculated using Eq. (4). The results are shown in Fig. 1.

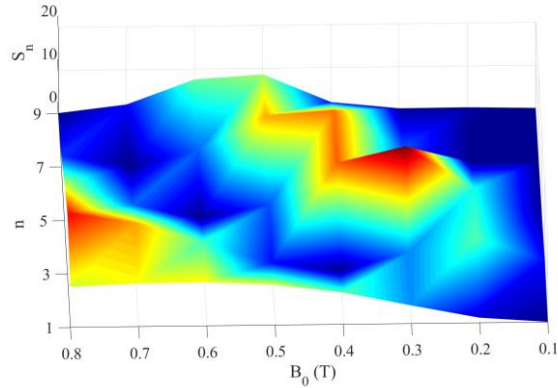

Figure 1. $S_n$ for different peak magnetic field of the subharmonic component.

It is found that the 5th or 7th harmonic radiation are enhanced when $B_0$ is in the range of 0.3–0.5 T. It should be noted that the fundamental harmonic radiation corresponds to the undulator with a period length of 24 mm. The 5th or 7th harmonic radiation corresponds to 5/3 or 7/3 harmonic of the 8mm periods length undulator.

The calculation above using a peak magnetic field of 0.68 T for the 8 mm period superconducting undulator is estimated from the scaling law. However, in a practical device, various constraints may reduce the achievable field. To check whether high harmonics can still be generated when lowering the magnetic field, we keep the biharmonic undulator periods length 8mm and 24mm and $d \equiv 2$ fixed, $B_0$ changes from 0.1T to 0.3 T, $S_n$ is calculated as shown in Fig. 2.

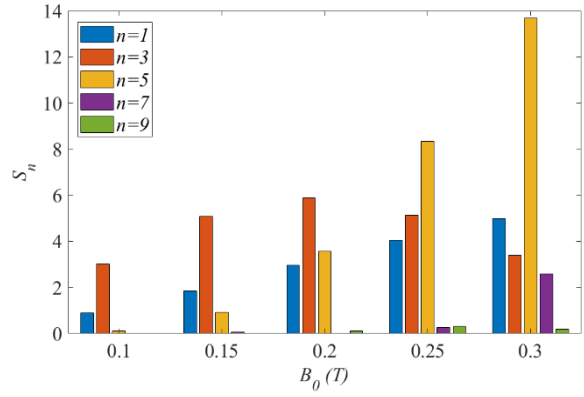

Figure 2. $S_n$ for different $B_0$ while keeping $h=3$ and $d=2$ fixed.

When the 5th harmonic radiation is greater than the 3rd, based on the shorter period field, the high harmonic radiation can be considered to be enhanced. From Fig. 2, it can be seen that when shorter period magnetic component decreases to 0.5 T, the high harmonic radiation can still be effectively generated.

## 3. Numerical results

In this section, we use SPECTRA [13] to calculate the on-axis flux and compare different cases with the biharmonic planar undulator to illustrate its properties as well as to verify whether the theoretical results presented in Section 1 are consistent with the numerical results. Here SSRF storage ring is taken for calculating [14], the parameters are listed in Table 1. The undulator total length is 1.2 m for various cases in this paper.

Table 1. Beam parameters for SSRF storage ring

| Parameter | Unit | Value |
| --- | --- | --- |
| Beam energy | GeV | 3.5 |
| Circumference | m | 432 |
| Natural emittance | nm rad | 3.9 |
| Energy spread | - | $9.7 \times 10^{-4}$ |
| Beam current | mA | 200 |
| Transverse coupling | - | 1% |
| Beta function @ straight section(x/y) | m | 3.6/2.5 |
| Dispersion @ straight section (x) | m | 0.1 |

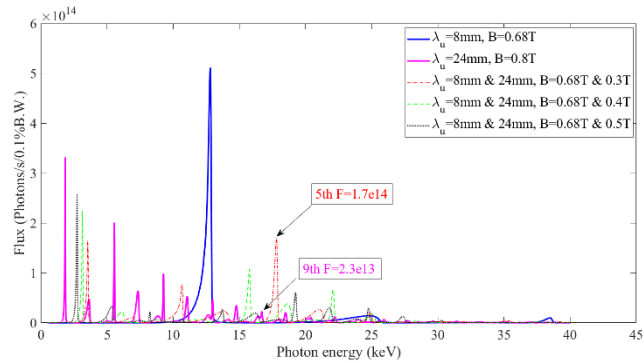

Figure 3. Flux calculated by SPECTRA.

The 'User Defined' field is input to SPECTRA. For the 'User Defined' field, only 'Near Field' can be calculated by SPECTRA. The energy dependence partial flux with a rectangular slit Δx=2mm, Δy=1mm is given out, as shown in Fig. 3. The numerical results are in good agreement with the theoretical predictions. For the biharmonic undulator with period lengths of 8 mm and 24 mm, and peak magnetic fields of 0.68 T and 0.3 T, respectively, the 5th harmonic radiation is significantly enhanced, as indicated by the red dashed line. For the case of subharmonic magnetic field of 0.4 T, the 3rd harmonic is strongly suppressed, while the 7th harmonic emerges which has been exactly indicated by Fig.1.

The biharmonic undulator exhibits a marked advantage in the photon energy range of 16–22 keV, corresponding to approximately 5/3-7/3 times the fundamental radiation photon energy of the shorter period undulator.

For comparison, the radiation from the superconducting undulator with a period length of 8 mm and a small $K \approx 0.5$ is calculated. As shown by the blue line in Fig. 3, high harmonics are difficult to generate, because of the small $K$. To cover 16-22 keV by the fundamental radiation is hopeless, since $K$ is already very small. From Eq. (9), it can be seen that even if the magnetic field is reduced to nearly zero, the radiation wavelength decreases by only about 10%.

In contrast, permanent-magnet undulators with a period length of, for example, 24 mm can reach $K \approx 2$, allowing the 7th, 9th, or even 11th harmonics to be readily produced. However, due to the relatively long period length, harmonics above the 9th order are required to reach photon energies around 16 keV. The flux of such high harmonics is significantly lower than that of the biharmonic undulator radiation, as shown by the magenta line in Fig. 3.

4. **Physical picture**

It is well known that undulator radiation is enhanced by the resonance effect between different periods [15]. Resonance occurs when the radiation from one period adds constructively in phase with that from other periods. The resonance wavelength is determined by the undulator period length and the beam energy as shown in Eq. (9).

$$\lambda_n = \frac{\lambda_u}{2n\gamma^2}\left(1 + \frac{1}{2}K^2 + \gamma^2\theta^2\right) \qquad (9)$$

Where $\theta$ is the observing angle apart from the beam axis. As the undulator period length decreases, the resonance wavelength is correspondingly reduced. If the magnetic field is relatively weak, the flux of bending magnet radiation around $\varepsilon_n$ (here $\varepsilon_n = \hbar\frac{2\pi c}{\lambda_n}$) is limited, the photon number for resonance is reduced as shown in Fig. 4, resulting a small flux of the high harmonic radiation. This picture is largely independent of the beam energy, since the critical photon energy of a bending magnet scales with the beam energy in the same manner which is proportional to $\gamma^2$.

$$\varepsilon_c(keV) = 0.665 E^2 (GeV^2) B(T) \qquad (10)$$

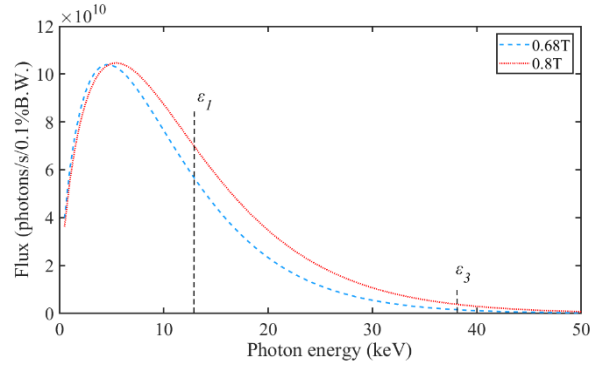

Figure 4. Flux of a bending magnet

From Eqs. (9) and (10), it shows that $\varepsilon_n \propto 1/\lambda_u$ and $\varepsilon_c \propto B$. The high harmonic content will depend on the undulator parameter $K$, which determines how many harmonics are covered by the bending magnet spectrum shown by Eq(11).

$$n \propto \varepsilon_c/\varepsilon_n \propto B\lambda_u \propto K \tag{11}$$

This highlights the challenge encountered when reducing the undulator period length, as achieving shorter magnetic period length requires even higher magnetic fields.

To increase the magnetic field, an additional field can be superimposed on the undulator. However, several conditions must be satisfied. First, both the first and second integrals of the magnetic field should vanish to ensure beam dynamic requirements; therefore, simply superimposing a uniform bending magnet field is not feasible. Second, the beam trajectory must exhibit a proper wiggling motion through the undulator. Consider the case of the $\lambda_u = 8$ mm(0.68 T) undulator field combined with a 24 mm(0.4 T) one. The corresponding total magnetic field is shown in Fig. 5(a). Between two adjacent magnetic field peaks, a field valley is required to sustain the beam oscillation, and this valley must fall to opposite sign so that the bending radius reversed as shown in Fig. 5(b). This explains why, as shown in Fig. 1, when the subharmonic undulator field exceeds 0.68 T, the high harmonic radiation diminishes, despite the increase in peak magnetic field, which may appear counterintuitive but reasonable.

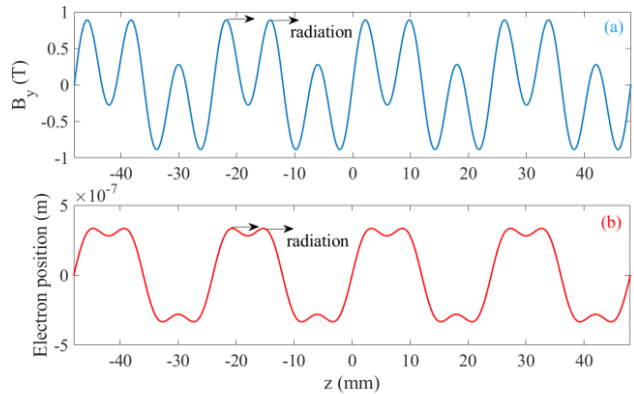

Figure 5. Field(a) and 2nd field integration of the field for a biharmonic undulator with 1/3 subharmonic.

With this physical picture in mind, a 1/2 subharmonic undulator can also be employed to generate a magnetic field similar to that shown in Fig. 6, provided that an appropriate phase shift is applied.

$$B_y = B_0[\sin(k_u z - \pi/4) + d\sin(2k_u z)] \tag{12}$$

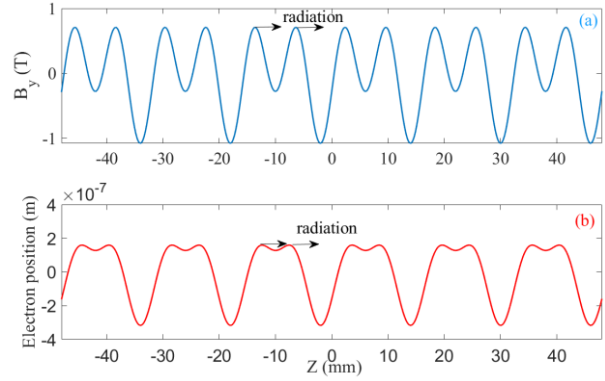

Figure 6. Field (a) and 2nd field integration of the field(b) for a biharmonic undulator with 1/2 subharmonic.

The theory of radiation property with initial magnetic field phase shift has not been included in Ref[11,12]. Thus we directly calculation the radiation with SPECTRA, with a 1/2 subharmonic ($\lambda_u$ = 16 mm) peak magnetic field of 0.4 T, the radiation properties are shown in Fig. 7.

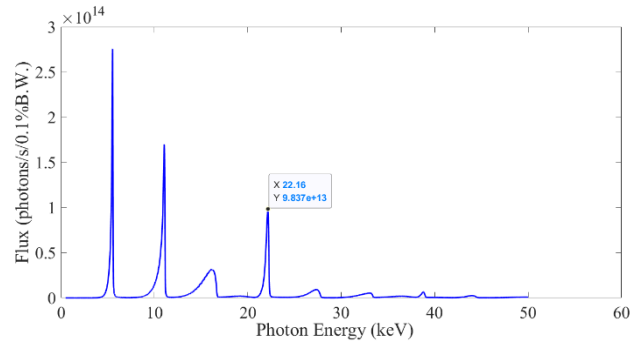

Figure 7. Flux calculated by SPECTRA for superimposing a 1/2 Subharmonic field.

It is found that the 4th harmonic at 22.16 keV is significantly enhanced. As pointed out in Ref. [9], for even harmonic field $h$, both even and odd harmonics can be enhanced on axis, which is confirmed by Fig. 7.

It is now clear that various subharmonic fields can be superimposed to form a biharmonic undulator, provided that the beam trajectory behaves as shown in Fig. 5 or Fig. 6. However, the 1/2 and 1/3 subharmonics are the most effective, as they correspond to the shortest period lengths.

## 5. Summary

This paper has demonstrated that a biharmonic planar undulator with a low-$K$ parameter can effectively enhance high harmonic radiation. The engineering feasibility is taken into account in selecting the undulator parameters in this study. A preliminary consideration of the undulator design is as shown in Fig.8. Following the concept of the APPLE-Knot undulator [16], two rows of auxiliary magnetic poles are introduced on both sides of the main poles, which will produce a subharmonic undulator field. As the field is not so high, the technical challenges remain manageable.

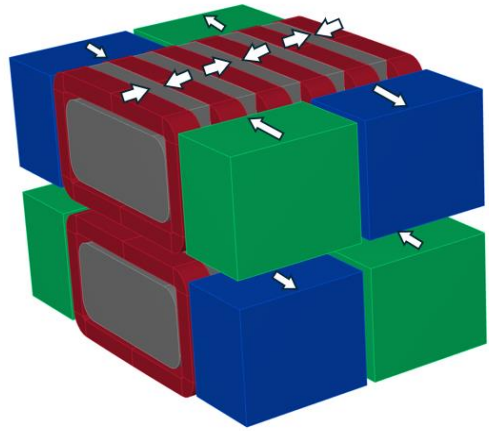

Figure 8. Sketch of a biharmonic undulator (the arrows denote magnetic field direction).

Compared with previous studies, we focus on the scenarios where $b > 1$, in which the high harmonic field is the dominant component, taking full advantage of the short-period undulator. The physical mechanism for high harmonic radiation generation under this configuration is presented in this paper and can be readily extended to other scenarios, such as the superposition of a 1/2 subharmonic magnetic field.

Before the advent of feasible superconducting undulators with $\lambda_u < 10 \text{ mm}$ and $K > 1$, the scheme proposed in this paper provides a viable solution for achieving unprecedented high flux radiation around the second harmonic of the short-period superconducting undulators.

**Acknowledgements**

This work was supported by the National Natural Science Foundation of China (Grant No. 12575157).

**Reference**

[1] S.H. Kim, A scaling law for the magnetic fields of superconducting undulators, Nuclear Instruments and Methods in Physics Research A **546** (2005) 604-619. https:/doi.org/10.1016/j.nima.2005.03.150

[2] K. Zhang and Marco Calvi, Review and prospects of world-wide superconducting undulator development for synchrotrons and FELs, Supercond. Sci. Technol. **35** (2022) 093001. https:/doi.org/10.1088/1361-6668/ac782a

[3] C. Benabderrahmane, et al., Development of PR$_2$FE$_{14}$B Cryogenic Undulator at Soleil, Proceedings of IPAC2011, San Sebastián, Spain, 3233-3235.

[4] M. Calvi, et al., GdBCO bulk superconducting helical undulator for x-ray free-electron lasers, Physical Review Research, **5** (2023) L032020. https:/doi.org/10.1103/PhysRevResearch.5.L032020

[5] S. Tantawi, et al., Experimental Demonstration of a Tunable Microwave Undulator, Physical Review Letters, **112** (2014) 164802. https:/doi.org/10.1103/PhysRevLett.112.164802

[6] J. E. Lawler, et al., Nearly copropagating sheared laser pulse FEL undulator for soft x-rays, Journal of Physics D: Applied Physics, **46** (2013) 325501 (11pp). https:/doi.org/10.1088/0022-3727/46/32/325501

[7] B. C. Jiang, et al., Using a Bessel light beam as an ultrashort period helical undulator, Physical Review Accelerators and Beams, **20** (2017) 070701. https:/doi.org/10.1103/PhysRevAccelBeams.20.070701

[8] K. Zhang et al., Record field in a 10 mm-period bulk high-temperature superconducting undulator, Supercond. Sci. Technol., vol. 36, no.5, Mar.2023, Art. no. 05LT01. https:/doi.org/10.1088/1361-6668/acc1a8

[9] G. Dattoli, V. V. Mikhailin, P. L. Ottaviani, K. V. Zhukovsky, Two-frequency Undulator and Harmonic Generation by an Ultra relativistic Electron, J. Appl. Phys. 100, 084507 (2006).

[10] K. V. Zhukovsky, Undulator and free-electron laser radiation with field harmonics and off-axis effects taken into account analytically, Physics-Uspekhi 64 (3) 304-316 (2021). https:/doi.org/10.3367/UFNe.2020.06.038803

[11] G. Dattolia, A. Doriaa, L. Giannessia, P.L. Ottaviani, Bunching and exotic undulator configurations in SASE FELs, Nuclear Instruments and Methods in Physics Research A **507** (2003) 388-391.

[12] A.M. Kalitenko, K.V. Zhukovskii, Radiation from Elliptical Undulators with Magnetic Field Harmonics, *J. Exp. Theor. Phys*. **130**, 327–337 (2020).   https:/doi.org/10.1134/S106377612001015X

[13] T. Tanaka, Major upgrade of the synchrotron radiation calculation code SPECTRA, J. Synchrotron Radiation 28, 1267 (2021). https:/doi.org/10.1107/S1600577521004100

[14] Liu Gui-Min, et al., Lattice Design for SSRF Storage Ring, Chinese Physics C, 2006, 30(S1): 144-146.

[15] Alexander Wu Chao, (2020), Lectures on Accelerator Physics, World Scientific.

[16] Fuhao Ji, et al., Design and performance of the APPLE-Knot undulator, J. Synchrotron Radiation 22, 901–907 (2015). http://dx.doi.org/10.1107/S1600577515006062